\documentclass[%
mathleft,%
final,%
]{an}
\usepackage{graphicx}
\usepackage{times}
\usepackage{amsmath}   
\usepackage{natbib}   
\usepackage{amssymb}    
\newcommand{\aap}{A\&A}

\begin{document}

\Pagespan{1}{}
\Yearpublication{2014}%
\Yearsubmission{2013}%
\Month{1}%
\Volume{335}%
\Issue{1}%
\DOI{This.is/not.aDOI}%

\title{Combining optical spectroscopy and interferometry}

\author{Stefan Kraus\inst{1}
}
\titlerunning{Combining optical spectroscopy and spectro-interferometry}
\authorrunning{Stefan Kraus}
\institute{
  University of Exeter, Astrophysics Group, Physics Building,
  Stocker Road, Exeter, EX4 4QL, UK}

\received{September 2, 2013}
\accepted{November 5, 2013}


\abstract{
Modern optical spectrographs and optical interferometers
push the limits in the spectral and spatial regime, providing 
important new tools for the exploration of the universe.  
In this contribution I outline the complementary nature 
of spectroscopic \& interferometric observations and 
discuss different strategies for combining such data.
Most remarkable, the latest generation of ``spectro-interferometric''
instruments combine the milliarcsecond angular resolution achievable
with interferometry with spectral capabilities, enabling
direct constraints on the distribution, density, 
kinematics, and ionization structure of the
gas component in protoplanetary disks.
I will present some selected studies
from the field of star- \& planet formation and hot star research
in order to illustrate these fundamentally new observational opportunities.
}

\maketitle

\section{Introduction}

Since more than a century, spectroscopy is one of the most fundamental and 
universal tools in observational astronomy.
With their ever-increasing sensitivity, spectral resolution, and calibration accuracy,
the latest generation of optical spectrographs enables new applications, both in galactic
and extragalactic astronomy.
For instance, in the field of star- and planet formation, 
spectroscopic and spectrophotometric observations are indispensible to characterize the 
dust emission in protoplanetary disks (with low spectral resolution, but wide wavelength coverage), 
or to determine the accretion and mass-loss processes in these objects (with high spectral resolution
observations in diagnostic lines).

For many applications, the main challenge in the interpretation of spectral line observations 
is the lack of information about the spatial distribution of the line-emitting gas.
This lack of spatial information is typically compensated by introducing
model assumptions, which allow us to relate a certain physical scenario with
the measured spectrum.
Unfortunately, these models often include a large number of free parameters,
which can result in {\it parameter degeneracies}, where different parameter
combinations result in similar spectra.
Another problem are {\it model degeneracies}, where different classes of
physical models can not be distinguished using spectroscopic data.
In some cases, these model classes might correspond to fundamentally different physical scenarios.
For instance, in the case of the Herbig~Ae/Be stars, it is still debated whether
the Br$\gamma$ emission in these objects traces mass accretion
or mass outflow processes \citep[see discussion in][]{kra08b}.
These degeneracies become more and more severe, as the models
become more and more complex.

Fortunately, many of the aforementioned limitations and degeneracies can now be solved using
optical interferometers such as ESO's Very Large Telescope Interferometer (VLTI),
which combine the light from separate optical apertures in order to reach
an unprecedented, milliarcsecond (mas) angular resolution.
In this contribution, I would like to outline how 
spatially and spectrally resolved observations can be combined
in order to solve some of the aforementioned model ambiguities,
where I consider three approaches:
In Section~\ref{sec:compl}, I discuss some of the opportunities provided by 
coordinated spectroscopic and interferometric observations,
while Section~\ref{sec:spectrointerferometry} discusses the ultimate
combination of these techniques in {\it spectro-interferometry}.
Another interesting and resources-efficient approach for obtaining
some gas kinematical constraints is {\it spectro-astrometry},
which I will discuss in Section~\ref{sec:spectroastrometry}.

\section{Complementary constraints from spectroscopy \& interferometry}
\label{sec:compl}

\begin{figure*}
  \includegraphics[angle=0, width=\linewidth]{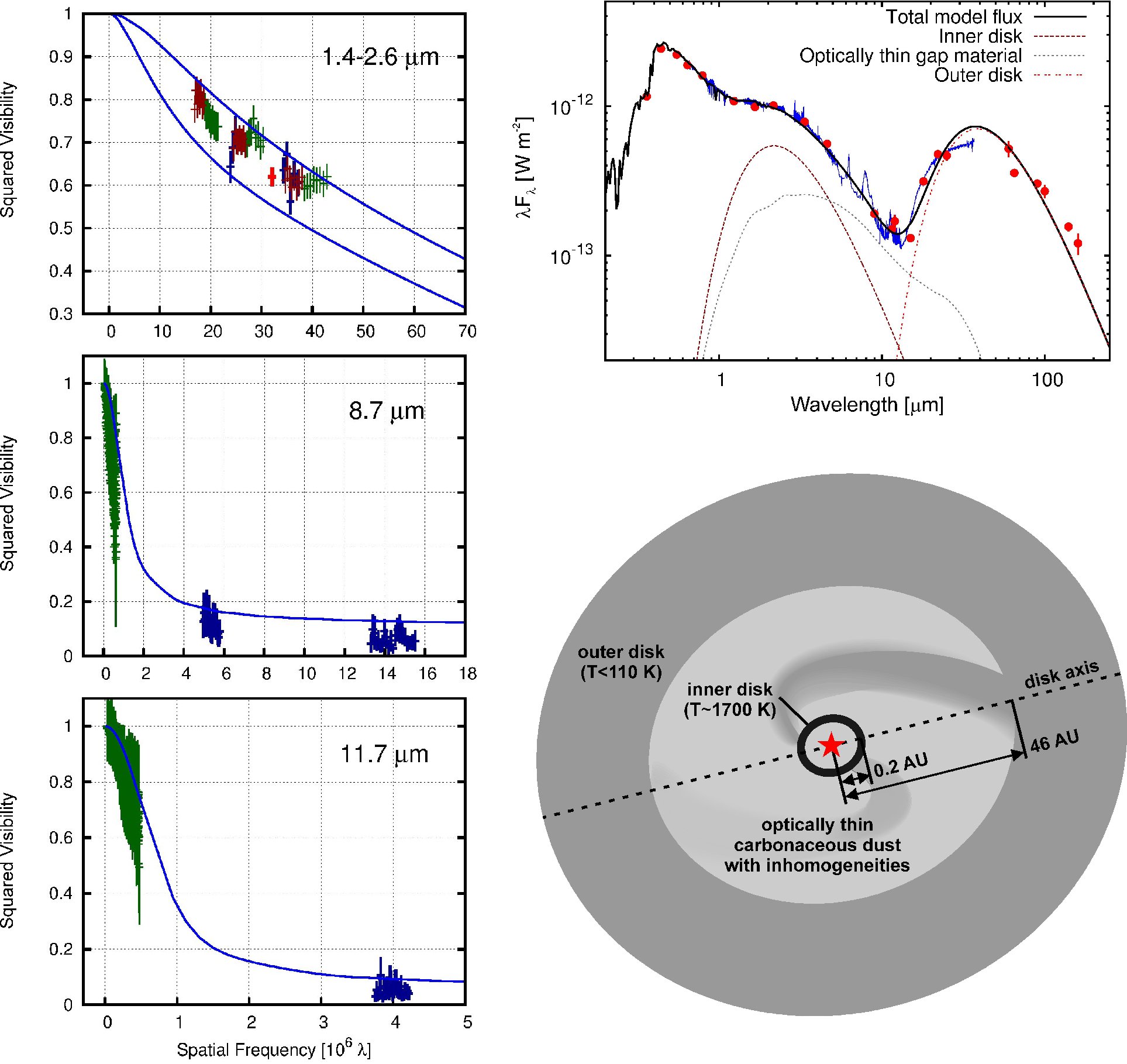}
  \caption{
    We use our multi-wavelength visibilities (left) and 
    spectral energy distribution (top right) to constrain
    the global geometry of the pre-transitional disk V1247\,Orionis.
    The sketch in the bottom right illustrates our
    best-fit model, which includes an extended gap that 
    separates the optically thick dust at the dust sublimation radius 
    from the outer disk at $\gtrsim 46$\,AU \citep{kra13}.}
  \label{fig:v1247ori}
\end{figure*}

\begin{figure}
  \includegraphics[angle=0, width=\linewidth]{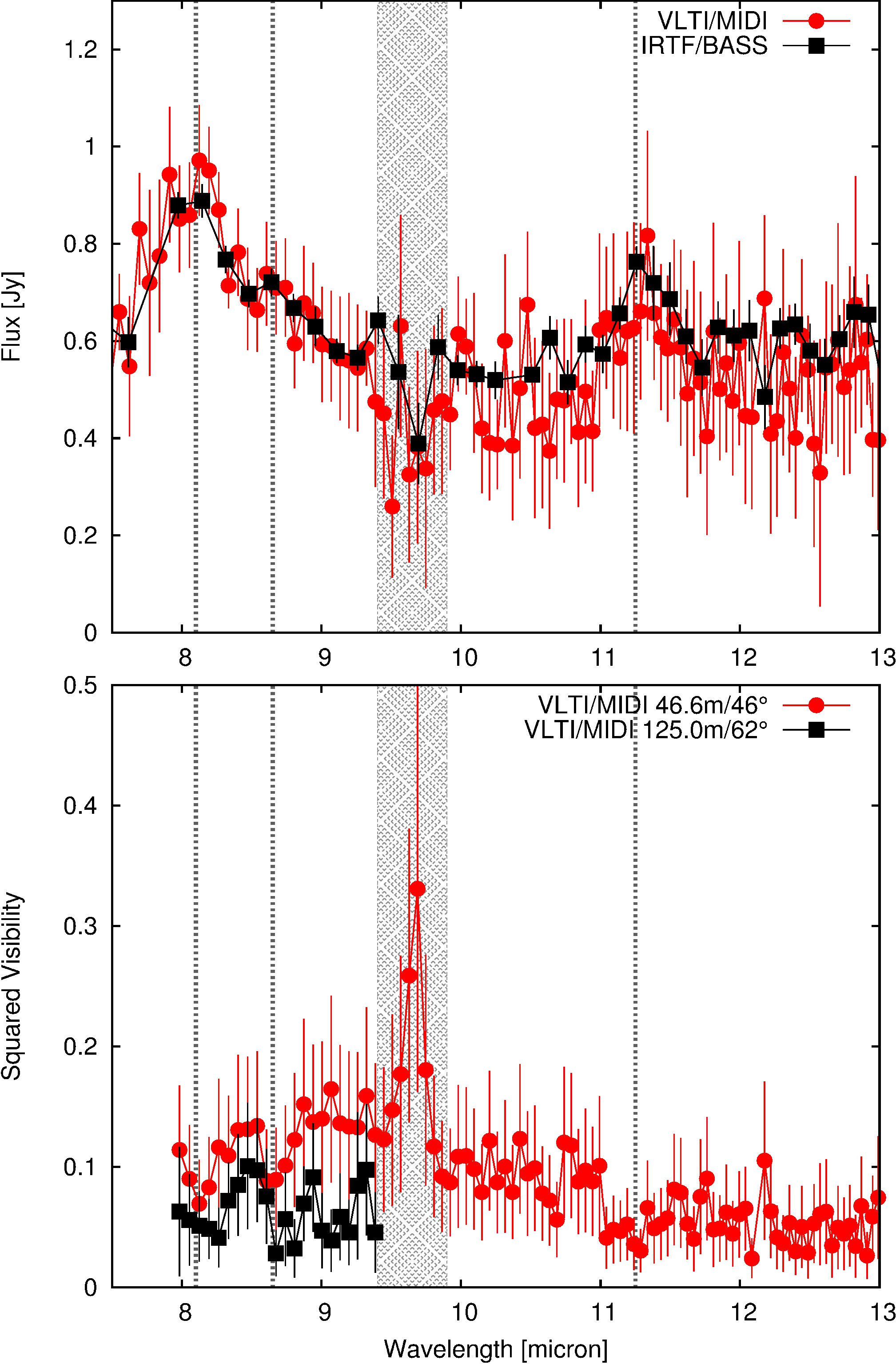}
  \caption{Spectrally dispersed mid-infrared visibilities measured on V1247\,Ori.
  The measurements in the gray shaded area are affected by an atmospheric Ozone absorption feature.
  Within the hydrocarbon (PAH) features (dashed lines), we detect a clear drop in visibility,
  indicating that these compounds are located at larger stellocentric radii than the
  continuum-emitting dust species. }
  \label{fig:PAH}
\end{figure}

In order to illustrate the complementary nature of spectroscopy and interferometry,
I discuss below our recent project on the pre-transitional disk of V1247\,Orionis \citep{kra13},
where we combined VLTI+Keck+Gemini infrared interferometry with IRTF, Spitzer, and HARPS spectroscopy.

V1247\,Ori is a F0-type star located in the Orion~OB1\,b association.
Earlier photometric observations on V1247\,Ori measured a strong near- and far-infrared excess,
but a deficit at mid-infrared wavelengths \citep{cab10}, compared to classical T~Tauri or
Herbig~Ae/Be disks.
Given the coarse sampling of these earlier photometric observations, we obtained new spectroscopic observations
using the near-infrared IRTF/SpeX (0.8-5.4\,$\mu$m) and mid-infrared IRTF/BASS (2.9-13.5\,$\mu$m) spectrograph,
and complement them with archival Spitzer/IRS (5.2-38\,$\mu$m) and HARPS ($V$-band) observations.
The resulting spectral energy distribution (SED, Fig.~\ref{fig:v1247ori}, top right) clearly 
shows that V1247\,Ori belongs to the class of ``pre-transitional disks'',
which are believed to exhibit extended disk gaps that might have been cleared by young planets
located in the gap region \citep{esp10}.
However, pure SED studies face serious ambiguities in solving the structure of these disks,
as not only the radial structure of these disks, but also the viewing geometry and dust composition is unknown.
Interferometry allows us to solve many of these ambiguities and to
determine the location of the emitting material directly.
For this purpose, we conducted near-infrared interferometric observations using 
the VLTI/AMBER and Keck Interferometer/V2-SPR instrument ($H$ and $K$-band),
as well as mid-infrared interferometry using the VLTI/MIDI instrument ($N$-band).
These long-baseline interferometric observations cover baseline lengths between 43 and 125\,m
and were complemented with Gemini/TreCS speckle interferometry and Keck/NIRC2 aperture masking interferometry,
which sample shorter baseline lengths $<8$\,m.

In order to model the interferometric data we require knowledge 
about the stellar properties. Unfortunately, the spectral type classification in the 
literature shows a wide spread from F0V to A5III.  Therefore, we repeated the classification using the HARPS
spectra and find best agreement with spectral type F0V ($T_{\rm eff}=7250$\,K),
where we adopt a distance of $385\pm 15$\,pc for the Alnilam cluster \citep{cab08b}.

Using our mid-infrared interferometric observations, we determined the disk inclination angle to $31.3 \pm 7.5^{\circ}$.
We then model the interferometric visibilities and SED data using a temperature power-law disk model (Fig.~\ref{fig:v1247ori}).
The near-infrared emission traces a ring-like inner disk that is located at the 
dust sublimation radius at 0.2\,AU. 
This narrow inner disk is separated from the outer disk that starts at 46\,AU.
Surprisingly, we find that the gap region is filled with significant amounts of optically 
thin material with a carbon-dominated dust mineralogy.
State-of-the-art SED models often associate this 
MIR excess emission to an heated wall of the outer disk \citep[e.g.][]{esp11}, 
which illustrates the importance of our multi-wavelength interferometry approach 
for unveiling the real geometry of these important objects. 

Besides the dust continuum, our spectrally dispersed VLTI/MIDI $N$-band observations also cover
spectral features corresponding to the hydrocarbon (PAH) emission (Fig.~\ref{fig:PAH}, top).
The measured visibilities show a small drop of visibility at the position of the PAH lines (Fig.~\ref{fig:PAH}, bottom), 
which indicates that the PAH-emission originates from significantly larger stellocentric radii 
than the dust continuum, likely the outer disk.

Finally, our Keck/NIRC2 aperture masking interferometry ($H$, $K'$, and $L$ band) reveals
non-zero phase signals, which indicate asymmetries in the brightness distribution 
on scales of 10..20\,AU, 
i.e.\ within the gap region. The detected asymmetries are highly significant, 
yet their amplitude and direction changes with wavelength, which is not consistent with a 
companion interpretation but indicates asymmetries in the disk structure itself.
We interpret this as strong evidence for the presence of complex density structures, 
possibly reflecting the dynamical interaction of the disk material with sub-stellar mass 
bodies that are responsible for the gap clearing.

V1247\,Ori exhibits photometric occultation events and spectroscopic variability \citep{cab10,kra13},
possibly indicating highly dynamic processes that might be at work in the inner disk regions.
An important objective of future investigations will be to
investigate the origin of this variability.
\citet{esp11} monitored a sample of pre-transitional disks
with Spitzer spectroscopy and found that the variability shows an anti-correlated 
behaviour at NIR and MIR wavelengths. 
In order to explain both the timescale and spectral behaviour of the variability, 
they proposed shadowing effects from co-rotating disk warps at the inner 
dust rim, probably triggered by orbiting planets. 
Such warps are also predicted by hydrodynamic simulations 
of disks with embedded planets \citep[e.g.][]{fou10}
and would result in a highly asymmetric brightness distribution.

Multi-epoch interferometric imaging observations could allow us to detect these
planet-induced disk structures and to trace their evolution and orbital motion 
around the star, which would provide new observational constraints on theories of 
disk dispersal and planet formation.
These interferometric observations should be accompanied with
extensive spectroscopic monitoring, which might enable to relate any
detected structural changes with the observed spectro-/photometric variability.

\section{Spectro-interferometry: Spectroscopy meets long-baseline interferometry}
\label{sec:spectrointerferometry}

\begin{figure}
  \centering
  \includegraphics[angle=0, width=7.45cm]{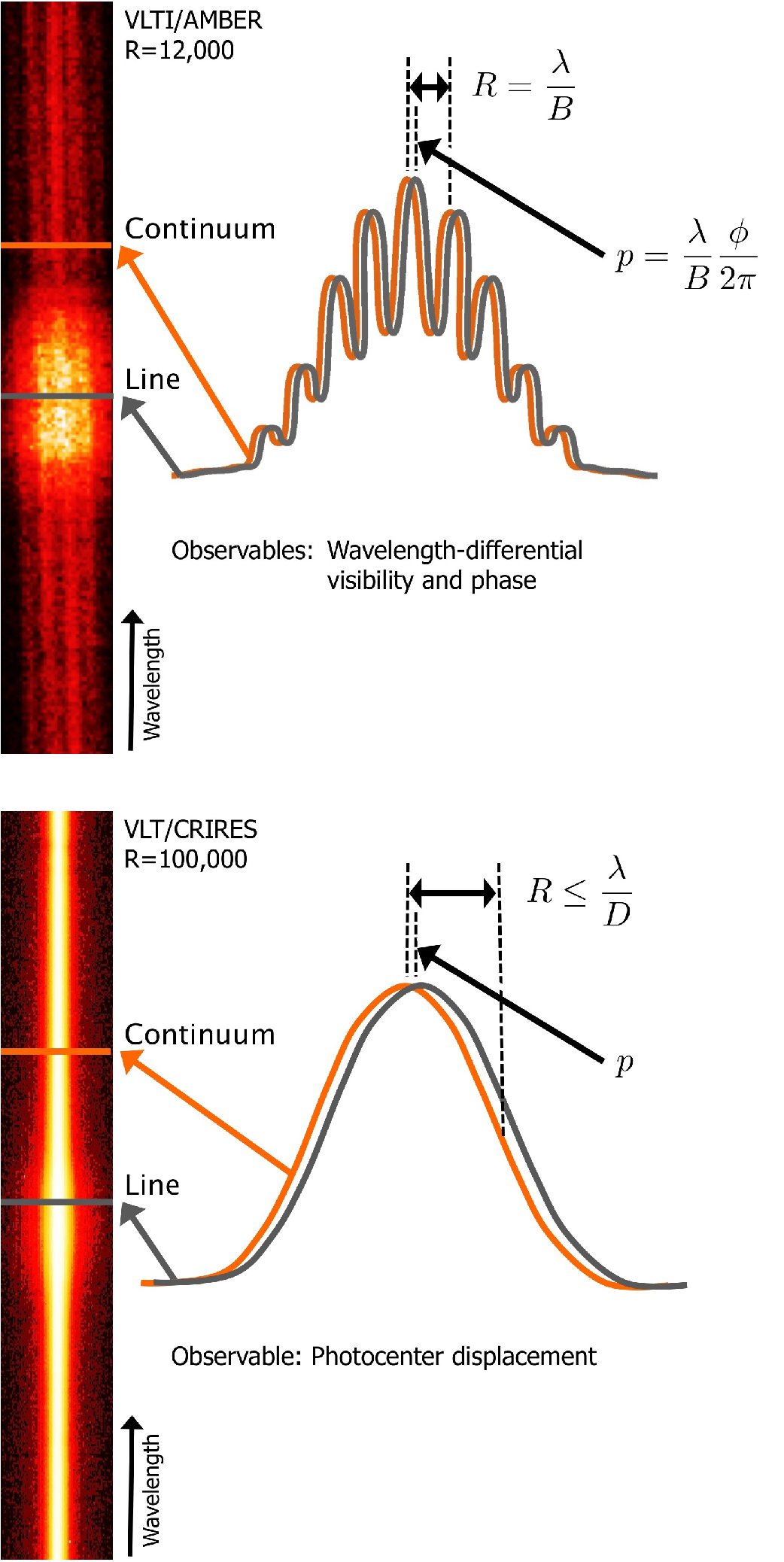}
  \caption{{\it Top:} Spectro-interferometry uses spectrally dispersed interferograms
    (such as the VLTI/AMBER $R=12,000$ interferogram from \citealt{wei07} shown here)
    in order to detect phase differences between the continuum emission and the emission
    within a spectral line.  In first-order expansion, this differential phase $\phi$
    correspond to the photocenter offset $p$ between the continuum- and line-emitting region.
    {\it Bottom:} 
    Spectro-astrometry computes the centroid position in conventional high-SNR spectra
    (such as the VLT/CRIRES spectrum from \citealt{kra12c} shown here),
    and measures the minuscule photocenter differences in the centroid position between the 
    continuum and line spectral channels.}
  \label{fig:comparison}
\end{figure}

In the last chapter, I outlined the advantages of coordinating
spectroscopic and interferometric observations in a joint observing campaign.
Even more powerful constraints are provided by {\it spectro-interferometric} instruments 
that incorporate high-resolution spectrographs in order to spectrally disperse the 
interferometric signals directly.  

From these dispersed interferograms, we can derive the visibility amplitude (interferogram contrast)
and phase (interferogram shift) in hundreds of spectral channels.
The visibilities and phases allow us to constrain the brightness distribution in each of these
spectral channels, covering simultaneously the continuum and emission or absorption lines.
A particularly useful quantity is the differential phase ($\phi$ in Fig.~\ref{fig:comparison}, top),
i.e.\ the phase difference between different spectral channels, which is also one of the few
quantities that are invariant to atmospheric perturbations \citep{pet89}.
For marginally resolved objects, the differential phase in a certain spectral line channel
is directly related to the photocenter displacement $p$ of the line-emitting region with respect
to the photocenter of the continuum emission, as projected on the telescope baseline.
The typical differential phase accuracy that can currently be reached is 
$\sim 1^{\circ}$, which for a 100\,m baseline corresponds to a photocenter displacement 
of just 0.012\,mas or 0.002\,AU ($=0.5\,R_{\sun}$) at the distance of the Taurus star forming region.
Measurements towards at least two different positon angles on the sky 
(which can be obtained from a single 3-telescope observations) 
are required in order to ``triangulate'' the 2-D photocenter vector 
$\overset{\rightarrow}{p}$ from these baseline-projected photocenter offsets.

The two operational instruments that offer this observing mode with sufficient
resolution for detailed kinematical studies are the near-infrared VLTI/AMBER 
instrument \citep[spectral resolution up to $\lambda/\Delta\lambda = 12\,000$,][]{pet07} and the
visual wavelength VEGA instrument at the CHARA array \citep[up to $\lambda/\Delta\lambda = 35\,000$,][]{mou12}.

\begin{figure*}
  \centering
  \includegraphics[angle=0, height=150mm]{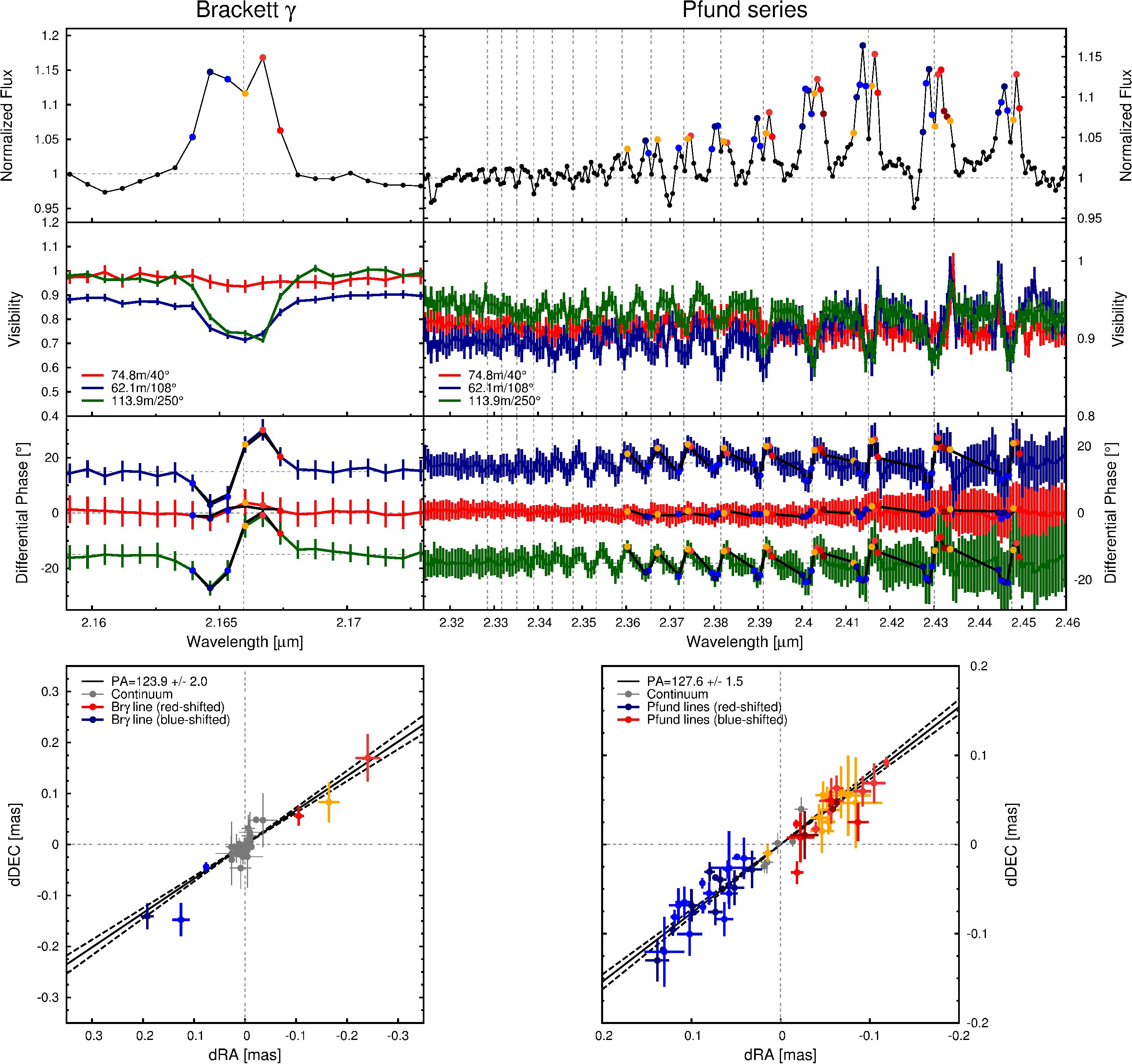}
  \caption{VLTI/AMBER spectro-interferometric data recorded on the classical Be star 
    $\zeta$\,Tau \citep{kra12a}, including spectra (top row), visibility amplitudes (2nd row from top),
    and differential phases (3rd row).
    A single spectral setup covers the Br$\gamma$-line (left), as well as several line transitions from the
    Pfund series (right), where each line profile is clearly double-peaked.
    We derived the photocenter displacement vector (bottom panels), which shows that the blue- and red-shifted emission
    originates from opposite directions of the stellar continuum emission, revealing the
    disk rotation axis. }
  \label{fig:ZTau}
\end{figure*}

For instance, in a recent study we used AMBER in order to study the disks around classical Be stars \citep{kra12a}.
Classical Be stars are surrounded by a gaseous disk that is formed of ejected
stellar material. However, both the ejection mechanism and the subsequent evolution 
of the disk material remain poorly understood. The proposed scenarios include 
radiatively driven winds, ram pressure or magnetically induced wind compression, 
and viscous decretion \citep{por03}.

\begin{figure}
  \centering
  \includegraphics[angle=0, width=72mm]{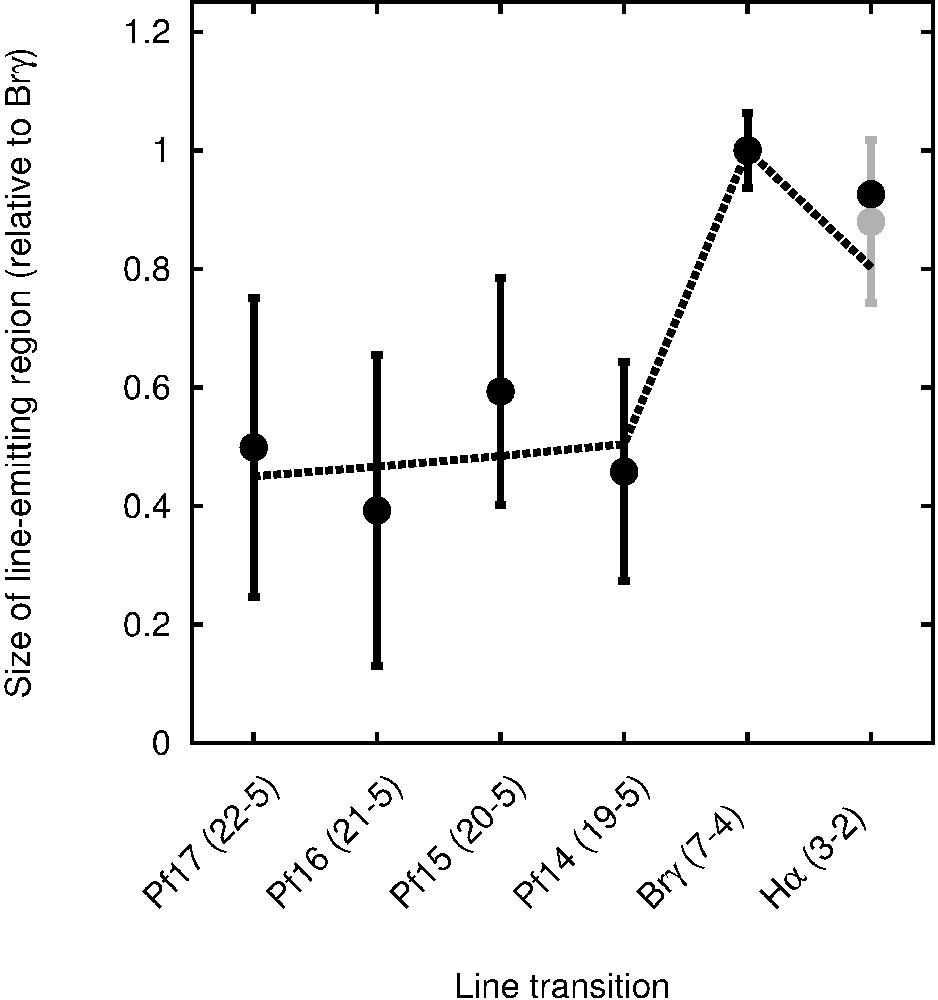}
  \caption{The size of the line-emitting region in the classical Be star $\zeta$\,Tau in
    different line transitions, including four Pfund lines, Br$\gamma$, and H$\alpha$ (measurements from
    \citealt{kra12a,qui94,vak98} and \citealt{tyc04}).
    The measured size differences can be reproduced with our simple LTE radiative transfer model (dashed line).
  }
  \label{fig:ZTauSize}
\end{figure}

Our study on the classical Be star $\zeta$\,Tau used AMBER's medium resolution code
($R=1500$) and covered the Br$\gamma$ 2.16\,$\mu$m and 
at least nine transitions from the Pfund series (Pf14-22, 2.359-2.448\,$\mu$m).
These lines can be covered with a single spectral setup in about one hour of
total observing time, which makes these observations highly efficient.
Both the Brackett and Pfund lines show a double-peaked profile (Fig.~\ref{fig:ZTau}, top row),
which is consistent with a disk rotation interpretation.
In Figure~\ref{fig:ZTau} (bottom panels), we show the photocenter displacements that we have
computed for $\zeta$\,Tau, both for the Br$\gamma$ line (left panel), and the Pf14-22 lines (right panel).
We find that the blue- and red-shifted line emission emerges from opposite directions 
with respect to the star, revealing the disk rotation axis (position angle $\sim 126^{\circ}$).
The measured visibilities show a significant drop, which indicates that the 
line emission originates from larger stellocentric radii than the underlying continuum emission.

We supplement our data with archival measurements 
for the H$\alpha$ line \citep{qui94,vak98,tyc04} and
find that the Br$\gamma$-line originates from similar stellocentric radii
as the H$\alpha$-line, while the Pfund lines originate much closer to the star.
Using a simple LTE model, we can reproduce the differences in the
spatial origin of the line transitions and constrain the 
temperature and excitation structure of the disk (Fig.~\ref{fig:ZTauSize}).
More detailed observations should enable us to derive also the
disk ionization and vertical temperature structure, which are currently
difficult to access.

Further improvements in the efficiency of uv-sampling will also soon enable 
the reconstruction of interferometric images for each velocity channel,
providing the equivalent to the ``channel maps'' in radio interferometry.
First attempts in this direction have been presented for instance 
by \citet{sch09b} and \citet{mil11}.

\section{Spectro-astrometry}
\label{sec:spectroastrometry}

\begin{figure*}[htb]
  \centering
  \includegraphics[angle=0, height=80mm]{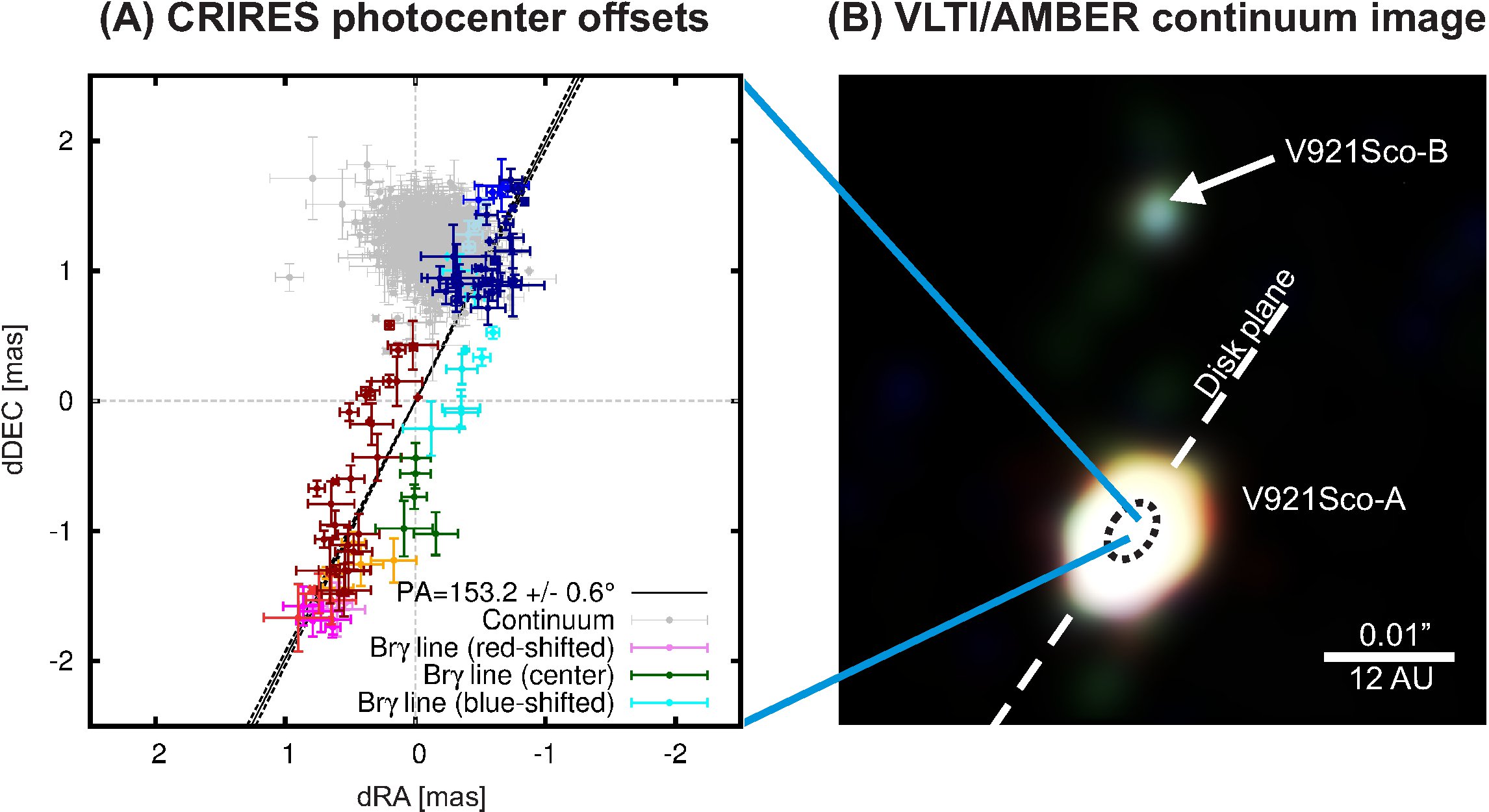}
  \caption{
    {\it Left:}
    Photocenter displacement vectors derived from our VLT/CRIRES spectra on V921\,Sco \citep{kra12c}.
    {\it Right:} 
    Color-composite of model-independent interferometric images \citep{kra12b}
    reconstructed from our low-spectral dispersion AMBER data on V921\,Sco 
    (blue: 1.65\,$\mu$m, green: 2.0\,$\mu$m, red: 2.3\,$\mu$m).
    The images have an effective resolution {$\lambda/2B=1.2$\,mas} and reveal 
    a previously unknown companion at a separation of 24.9\,mas (corresponding to 29\,AU at 1.15\,kpc).
  }
  \label{fig:V921Sco1}
\end{figure*}

\begin{figure*}
  \centering
  \includegraphics[angle=0, height=125mm]{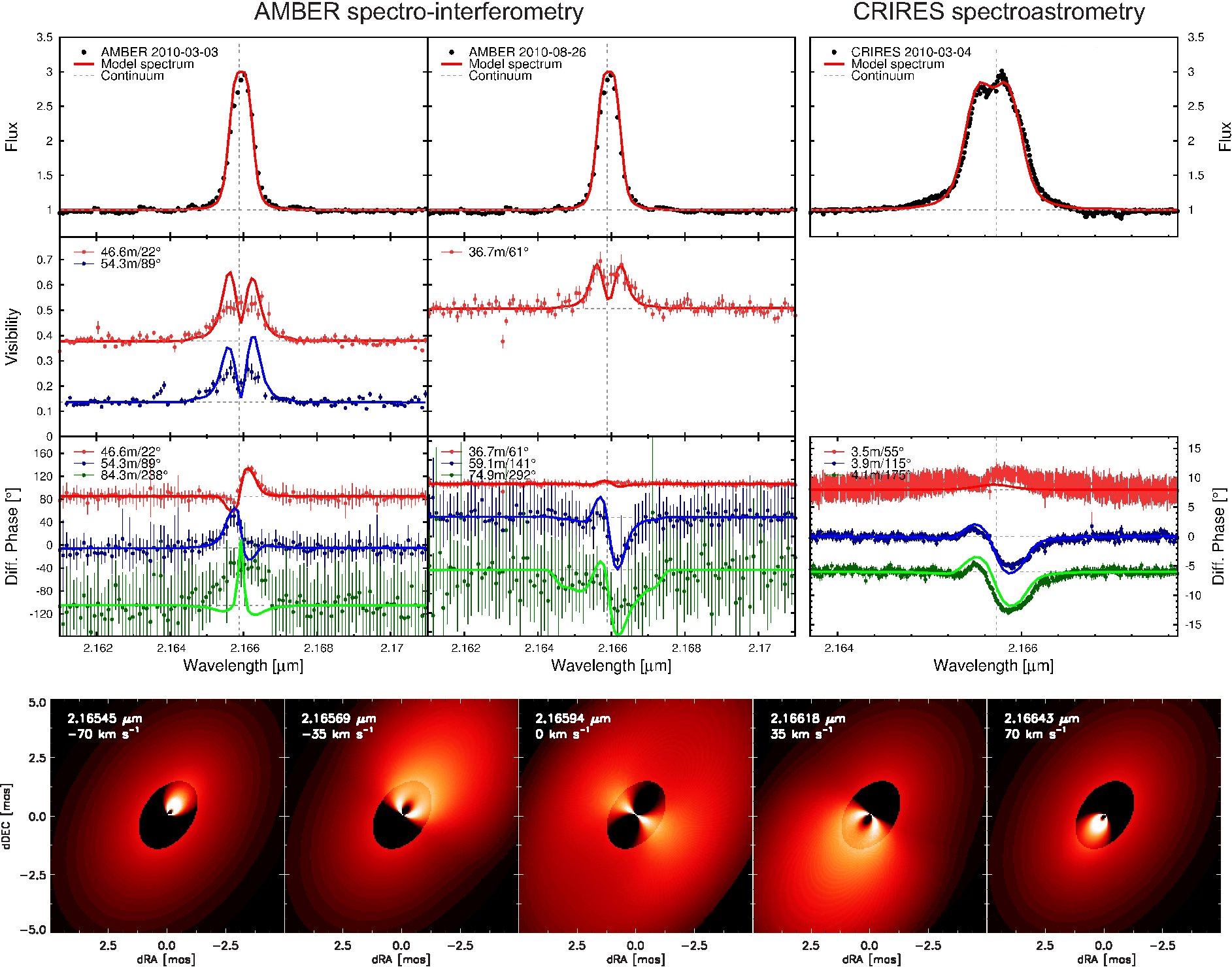}
  \caption{
    {\it Top:} 
    To constrain our V921\,Sco model, we fitted the VLTI/AMBER spectra, visibilities, and differential phases (left and middle row) 
    simultaneously with the VLT/CRIRES spectra and spectro-astrometric signal (right row).
    {\it Bottom:}
    Our best-fit kinematical model consists of a Keplerian-rotating disk that emits in Br$\gamma$
    at stellocentric radii from a few stellar radii ($<0.5$\,AU) to $\sim 6.5$\,AU, while 
    the dust disk starts at 1.8\,AU \citep{kra12c}.
    }
  \label{fig:V921Sco2}
\end{figure*}

Another technique that can provide detailed kinematical and spatial information about
the gas distribution on AU-scales is spectro-astrometry.
Spectro-astrometry uses high-SNR long-slit spectra to measure the
centroid position of an apparently unresolved object as function of wavelength
(Fig.~\ref{fig:comparison}, bottom).
By design, this technique is not able to resolve the line-emitting region directly 
(e.g.\ it does not provide visibility information), but it can reveal small-scale 
photocenter displacements, which are mathematically equivalent to the 
wavelength-differential phases measured in spectro-interferometry in the short-baseline regime.
This makes it possible to combine the AMBER (visibility, differential phase, closure phase) 
and the spectro-astrometric signals
directly for quantitative modeling, providing highly complementary constraints.

For our combined spectro-interferometry + spectro-astrometric observations, 
we employed the VLTI/AMBER ($R=12,000$) and VLT/CRIRES ($R=100,000$) instruments
and targeted the B[e] star V921\,Scorpii \citep{kra12b,kra12c}.
Important aspects about V921\,Sco are still strongly debated, including its distance and 
evolutionary stage, where both a pre-main-sequence \citep[e.g.][]{ben98}
and a post-main-sequence \citep[e.g.][]{bor07}
nature has been suggested.

The CRIRES spectra were obtained with the slit oriented towards three position angles (55, 115, 175$^{\circ}$)
and the corresponding anti-parallel positions (235, 295, 355$^{\circ}$).
The spectro-astrometric signals were derived for each position angle separately
and then the signals towards parallel and anti-parallel position angles were subtracted, 
which provides an efficient method for eliminating instrumental artefacts \citep{bra06}.

Considering the CRIRES spectro-astrometric measurements alone leads to puzzling results:
The photocenter in the red-shifted line channels (red data points in Fig.~\ref{fig:V921Sco1}, left)
shows a significant offset of $\sim 3$\,mas, while the blue-shifted channels 
(blue data points) do not show a significant displacement with respect to the
continuum channels (grey data points).
This behaviour is not consistent with the expected signature for a rotating circumstellar disk,
where the photocenter of the blue and red-shifted line emission should be displaced 
with similar amplitude in opposite directions from the central star,
but might guide us to the interpretation that the Br$\gamma$ emission is associated,
for instance, with a one-sided jet instead of a rotation disk.

However, the puzzle was solved with a VLTI/AMBER aperture synthesis image that 
we reconstructed from AMBER continuum data ($H$ and $K$-band; Fig.~\ref{fig:V921Sco1}, right).
The image reveals a previously unknown companion that is located at a separation of $25.0 \pm 0.8$\,mas ($\sim 29$\,AU)
in northern direction (PA $353.8 \pm 1.6^{\circ}$) from the disk-harboring B0V-type star.
Therefore, the continuum photocenter is shifted systematically towards the 
direction of the companion, while the primary star is located to a good accuracy 
half-way between the blue- and red-shifted photocenter vectors,
consistent with a disk scenario.

Taking the presence of the companion into account, we then engaged in a detailed
modeling of the spectro-astrometric and the detected strong AMBER visibility 
and differential phase signals (Fig.~\ref{fig:V921Sco2}, 2nd and 3rd row from top).
For this purpose, we tested rotation disks with different velocity profiles and
found reasonable agreement with a Keplerian disk, where most of the line-emitting gas 
is located inside of the dust sublimation radius that we constrain to $1.59 \pm 0.25$\,mas 
or $\sim 1.8$\,AU (see best-fit model images in Fig.~\ref{fig:V921Sco2}, bottom).

Our detection of a Keplerian velocity field 
provides important evidence for the pre-main-sequence (Herbig~B[e]) nature of V921\,Sco,
as the decretion disks around post-main-sequence supergiant B[e] stars 
are expected to exhibit a strong outflowing velocity component \citep{lam91},
which is not observed in our data.  Also, we find that the previous distance estimates
have likely underestimated the distance to V921\,Sco ($d=1.15$\,kpc), 
as this distance implies with our measured rotation profile a stellar mass 
of just $5.4 \pm 0.4\,M_{\sun}$, which
is significantly lower than expected from the spectral classification.

\section{Conclusions}
\label{sec:conclusions}

Spectroscopy and interferometry provide highly complementary constraints
and extend our observational capabilities in the spectral and spatial domain.
We are just starting to explore the scientific opportunities that arise 
from combining these techniques, but it is clear that the unique combination
of high spatial and high spectral resolution could enable transformational studies
both in galactic and extragalactic astronomy.

Particularly powerful constraints can be obtained with spectro-interferometry,
as I have illustrated with our studies on $\zeta$\,Tau and V921\,Sco,
where we constrained the spatial distribution and kinematics of the
line-emitting hydrogen gas on scales of a few stellar radii.
For faint ($K \lesssim 8$) or strongly resolved objects, it is still rather challenging 
to perform these observations, mainly due to the stringent requirements on fringe
tracking and the low efficiency of filling the $uv$-plane with 3-telescope interferometric observations.
Therefore, we investigated the possibility of combining spectro-interferometry
with spectro-astrometric observations that can be conducted with conventional spectrographs,
providing a resource-efficient approach to obtain indispensable constraints for a kinematical modeling.

Efficient model-independent imaging should become accessible with the upcoming generation
of spectro-interferometric 4-telescope beam combiners and will enable detailed studies on
complex velocity structures or on time-variable processes.


%

\begin{thebibliography}{24}
\expandafter\ifx\csname natexlab\endcsname\relax\def\natexlab#1{#1}\fi

\bibitem[{{Benedettini} {et~al.}(1998){Benedettini}, {Nisini}, {Giannini},
  {Lorenzetti}, {Tommasi}, {Saraceno}, \& {Smith}}]{ben98}
{Benedettini}, M., {Nisini}, B., {Giannini}, T., {Lorenzetti}, D., {Tommasi},
  E., {Saraceno}, P., \& {Smith}, H.~A. 1998, \aap, 339, 159

\bibitem[{{Borges Fernandes} {et~al.}(2007){Borges Fernandes}, {Kraus}, {Lorenz
  Martins}, \& {de Ara{\'u}jo}}]{bor07}
{Borges Fernandes}, M., {Kraus}, M., {Lorenz Martins}, S., \& {de Ara{\'u}jo},
  F.~X. 2007, \mnras, 377, 1343

\bibitem[{{Brannigan} {et~al.}(2006){Brannigan}, {Takami}, {Chrysostomou}, \&
  {Bailey}}]{bra06}
{Brannigan}, E., {Takami}, M., {Chrysostomou}, A., \& {Bailey}, J. 2006,
  \mnras, 367, 315

\bibitem[{{Caballero}(2010)}]{cab10}
{Caballero}, J.~A. 2010, \aap, 511, L9

\bibitem[{{Caballero} \& {Solano}(2008)}]{cab08b}
{Caballero}, J.~A., \& {Solano}, E. 2008, \aap, 485, 931

\bibitem[{{Espaillat} {et~al.}(2011){Espaillat}, {Furlan}, {D'Alessio},
  {Sargent}, {Nagel}, {Calvet}, {Watson}, \& {Muzerolle}}]{esp11}
{Espaillat}, C., {Furlan}, E., {D'Alessio}, P., {Sargent}, B., {Nagel}, E.,
  {Calvet}, N., {Watson}, D.~M., \& {Muzerolle}, J. 2011, \apj, 728, 49

\bibitem[{{Espaillat} {et~al.}(2010){Espaillat}, {D'Alessio}, {Hern{\'a}ndez},
  {Nagel}, {Luhman}, {Watson}, {Calvet}, {Muzerolle}, \& {McClure}}]{esp10}
{Espaillat}, C., {et~al.} 2010, \apj, 717, 441

\bibitem[{{Fouchet} {et~al.}(2010){Fouchet}, {Gonzalez}, \& {Maddison}}]{fou10}
{Fouchet}, L., {Gonzalez}, J.-F., \& {Maddison}, S.~T. 2010, \aap, 518, A16

\bibitem[{{Kraus} {et~al.}(2012{\natexlab{a}}){Kraus}, {Calvet}, {Hartmann},
  {Hofmann}, {Kreplin}, {Monnier}, \& {Weigelt}}]{kra12b}
{Kraus}, S., {Calvet}, N., {Hartmann}, L., {Hofmann}, K.-H., {Kreplin}, A.,
  {Monnier}, J.~D., \& {Weigelt}, G. 2012{\natexlab{a}}, \apjl, 746, L2

\bibitem[{{Kraus} {et~al.}(2012{\natexlab{b}}){Kraus}, {Calvet}, {Hartmann},
  {Hofmann}, {Kreplin}, {Monnier}, \& {Weigelt}}]{kra12c}
---. 2012{\natexlab{b}}, \apj, 752, 11

\bibitem[{{Kraus} {et~al.}(2008){Kraus}, {Hofmann}, {Benisty}, {Berger},
  {Chesneau}, {Isella}, {Malbet}, {Meilland}, {Nardetto}, {Natta}, {Preibisch},
  {Schertl}, {Smith}, {Stee}, {Tatulli}, {Testi}, \& {Weigelt}}]{kra08b}
{Kraus}, S., {et~al.} 2008, \aap, 489, 1157

\bibitem[{{Kraus} {et~al.}(2012{\natexlab{c}}){Kraus}, {Monnier}, {Che},
  {Schaefer}, {Touhami}, {Gies}, {Aufdenberg}, {Baron}, {Thureau}, {ten
  Brummelaar}, {McAlister}, {Turner}, {Sturmann}, \& {Sturmann}}]{kra12a}
---. 2012{\natexlab{c}}, \apj, 744, 19

\bibitem[{{Kraus} {et~al.}(2013){Kraus}, {Ireland}, {Sitko}, {Monnier},
  {Calvet}, {Espaillat}, {Grady}, {Harries}, {H{\"o}nig}, {Russell},
  {Swearingen}, {Werren}, \& {Wilner}}]{kra13}
---. 2013, \apj, 768, 80

\bibitem[{{Lamers} \& {Pauldrach}(1991)}]{lam91}
{Lamers}, H.~J.~G., \& {Pauldrach}, A.~W.~A. 1991, \aap, 244, L5

\bibitem[{{Millour} {et~al.}(2011){Millour}, {Meilland}, {Chesneau}, {Stee},
  {Kanaan}, {Petrov}, {Mourard}, \& {Kraus}}]{mil11}
{Millour}, F., {Meilland}, A., {Chesneau}, O., {Stee}, P., {Kanaan}, S.,
  {Petrov}, R., {Mourard}, D., \& {Kraus}, S. 2011, \aap, 526, A107+

\bibitem[{{Mourard} {et~al.}(2012){Mourard}, {Challouf}, {Ligi}, {B{\'e}rio},
  {Clausse}, {Gerakis}, {Bourges}, {Nardetto}, {Perraut}, {Tallon-Bosc},
  {McAlister}, {ten Brummelaar}, {Ridgway}, {Sturmann}, {Sturmann}, {Turner},
  {Farrington}, \& {Goldfinger}}]{mou12}
{Mourard}, D., {et~al.} 2012, in Society of Photo-Optical Instrumentation
  Engineers (SPIE) Conference Series, Vol. 8445, Society of Photo-Optical
  Instrumentation Engineers (SPIE) Conference Series

\bibitem[{{Petrov}(1989)}]{pet89}
{Petrov}, R.~G. 1989, in NATO ASIC Proc. 274: Diffraction-Limited Imaging with
  Very Large Telescopes, ed. D.~M. {Alloin} \& J.-M. {Mariotti}, 249

\bibitem[{{Petrov} {et~al.}(2007){Petrov}, {Malbet}, {Weigelt}, {Antonelli},
  {Beckmann}, {Bresson}, {Chelli}, {Dugu{\'e}}, {Duvert}, {Gennari},
  {Gl{\"u}ck}, {Kern}, {Lagarde}, {Le Coarer}, {Lisi}, {Millour}, {Perraut},
  {Puget}, {Rantakyr{\"o}}, {Robbe-Dubois}, {Roussel}, {Salinari}, {Tatulli},
  {Zins}, {Accardo}, {Acke}, {Agabi}, {Altariba}, {Arezki}, {Aristidi},
  {Baffa}, {Behrend}, {Bl{\"o}cker}, {Bonhomme}, {Busoni}, {Cassaing},
  {Clausse}, {Colin}, {Connot}, {Delboulb{\'e}}, {Domiciano de Souza},
  {Driebe}, {Feautrier}, {Ferruzzi}, {Forveille}, {Fossat}, {Foy},
  {Fraix-Burnet}, {Gallardo}, {Giani}, {Gil}, {Glentzlin}, {Heiden},
  {Heininger}, {Hernandez Utrera}, {Hofmann}, {Kamm}, {Kiekebusch}, {Kraus},
  {Le Contel}, {Le Contel}, {Lesourd}, {Lopez}, {Lopez}, {Magnard}, {Marconi},
  {Mars}, {Martinot-Lagarde}, {Mathias}, {M{\`e}ge}, {Monin}, {Mouillet},
  {Mourard}, {Nussbaum}, {Ohnaka}, {Pacheco}, {Perrier}, {Rabbia}, {Rebattu},
  {Reynaud}, {Richichi}, {Robini}, {Sacchettini}, {Schertl}, {Sch{\"o}ller},
  {Solscheid}, {Spang}, {Stee}, {Stefanini}, {Tallon}, {Tallon-Bosc}, {Tasso},
  {Testi}, {Vakili}, {von der L{\"u}he}, {Valtier}, {Vannier}, \&
  {Ventura}}]{pet07}
{Petrov}, R.~G., {et~al.} 2007, \aap, 464, 1

\bibitem[{{Porter} \& {Rivinius}(2003)}]{por03}
{Porter}, J.~M., \& {Rivinius}, T. 2003, \pasp, 115, 1153

\bibitem[{{Quirrenbach} {et~al.}(1994){Quirrenbach}, {Buscher}, {Mozurkewich},
  {Hummel}, \& {Armstrong}}]{qui94}
{Quirrenbach}, A., {Buscher}, D.~F., {Mozurkewich}, D., {Hummel}, C.~A., \&
  {Armstrong}, J.~T. 1994, \aap, 283, L13

\bibitem[{{Schmitt} {et~al.}(2009){Schmitt}, {Pauls}, {Tycner}, {Armstrong},
  {Zavala}, {Benson}, {Gilbreath}, {Hindsley}, {Hutter}, {Johnston},
  {Jorgensen}, \& {Mozurkewich}}]{sch09b}
{Schmitt}, H.~R., {et~al.} 2009, \apj, 691, 984

\bibitem[{{Tycner} {et~al.}(2004){Tycner}, {Hajian}, {Armstrong}, {Benson},
  {Gilbreath}, {Hutter}, {Lester}, {Mozurkewich}, \& {Pauls}}]{tyc04}
{Tycner}, C., {et~al.} 2004, \aj, 127, 1194

\bibitem[{{Vakili} {et~al.}(1998){Vakili}, {Mourard}, {Stee}, {Bonneau},
  {Berio}, {Chesneau}, {Thureau}, {Morand}, {Labeyrie}, \&
  {Tallon-Bosc}}]{vak98}
{Vakili}, F., {et~al.} 1998, \aap, 335, 261

\bibitem[{{Weigelt} {et~al.}(2007){Weigelt}, {Kraus}, {Driebe}, {Petrov},
  {Hofmann}, {Millour}, {Chesneau}, {Schertl}, {Malbet}, {Hillier}, {Gull},
  {Davidson}, {Domiciano de Souza}, {Antonelli}, {Beckmann}, {Bresson},
  {Chelli}, {Dugu{\'e}}, {Duvert}, {Gennari}, {Gl{\"u}ck}, {Kern}, {Lagarde},
  {Le Coarer}, {Lisi}, {Perraut}, {Puget}, {Rantakyr{\"o}}, {Robbe-Dubois},
  {Roussel}, {Tatulli}, {Zins}, {Accardo}, {Acke}, {Agabi}, {Altariba},
  {Arezki}, {Aristidi}, {Baffa}, {Behrend}, {Bl{\"o}cker}, {Bonhomme},
  {Busoni}, {Cassaing}, {Clausse}, {Colin}, {Connot}, {Delboulb{\'e}},
  {Feautrier}, {Ferruzzi}, {Forveille}, {Fossat}, {Foy}, {Fraix-Burnet},
  {Gallardo}, {Giani}, {Gil}, {Glentzlin}, {Heiden}, {Heininger}, {Hernandez
  Utrera}, {Kamm}, {Kiekebusch}, {Le Contel}, {Le Contel}, {Lesourd}, {Lopez},
  {Lopez}, {Magnard}, {Marconi}, {Mars}, {Martinot-Lagarde}, {Mathias},
  {M{\`e}ge}, {Monin}, {Mouillet}, {Mourard}, {Nussbaum}, {Ohnaka}, {Pacheco},
  {Perrier}, {Rabbia}, {Rebattu}, {Reynaud}, {Richichi}, {Robini},
  {Sacchettini}, {Sch{\"o}ller}, {Solscheid}, {Spang}, {Stee}, {Stefanini},
  {Tallon}, {Tallon-Bosc}, {Tasso}, {Testi}, {Vakili}, {von der L{\"u}he},
  {Valtier}, {Vannier}, {Ventura}, {Weis}, \& {Wittkowski}}]{wei07}
{Weigelt}, G., {et~al.} 2007, \aap, 464, 87

\end{thebibliography}
%

\end{document}